\documentclass[aps,amsmath,amssymb,prb,twocolumn,showpacs]{revtex4}
\usepackage{bm}
\usepackage{epsfig}

\begin{document}

\title{FLUX-LATTICE MELTING IN TWO-DIMENSIONAL DISORDERED SUPERCONDUCTORS}

\author{Mai Suan Li}
\affiliation{Institute of Physics, Al. Lotnikow 32/46, 02-668
Warsaw, Poland}

\author{Thomas Nattermann}
\affiliation{Institut f\"ur Theoretische Physik, Universit\"at zu
K\"oln, Z\"ulpicher Str. 77, D-50937 K\"oln, Germany}

\date{17.1.03}

\begin{abstract}
The flux line lattice melting transition in two-dimensional pure and
disordered superconductors
is studied by a Monte Carlo simulation using the lowest Landau level
approximation and quasi-periodic boundary condition on a plane.
The position of the melting line was determined from the diffraction pattern
of the superconducting order parameter. In the clean case we confirmed
the results from earlier studies which show
the existence of a quasi-long range ordered vortex lattice at low temperatures.
Adding frozen disorder to the system the melting transition line is shifted
to slightly lower fields.
The correlations of the order parameter for translational long range order
of the vortex positions seem to decay slightly faster than a power law 
(in agreement with the theory of Carpentier and Le Doussal) although a
simple power law decay cannot be excluded. The corresponding positional
glass correlation function decays as a power law establishing the existence
 of a quasi-long range ordered positional glass formed by the vortices.
The correlation function characterizing a phase coherent vortex glass decays
however exponentially ruling out the possible existence of a phase coherent
vortex glass phase.  
\end{abstract}

\vspace{1cm}


\maketitle

\section{Introduction}

The nature of the mixed phase in type-II two-dimensional (2D) superconductors,
despite considerable experimental, theoretical and numerical effort, remains
unclear. In the mean field approximation the vortices of a pure system
are known to form 
an Abrikosov lattice \cite{Abrikosov57}.
In D=3 dimensions thermal fluctuations reduce the upper critical field line as
first shown by Eilenberger \cite{Eilenberger67} leaving a vortex liquid 
phase between the transition line and the mean field 
$H_{c2}(T)$.
In D=2 dimensions
the phase diagram  between the critical field 
lines $H_{c1}(T)$ and  $H_{c2}(T)$ is
still under debate.
 Applying the
Kosterlitz-Thouless-Berezinskii-Halperin-Nelson-Young (KTBHNY) theory
\cite{Kosterlitz73,Berezinsky,Halperin78,Nelson79,Young79} originally 
developed for
melting in 2D crystals to superconductors \cite{Doniach79,Fisher80}
one obtains a phase diagram with a solid, a hexatic and a liquid phase.
The hexatic phase was observed in recent simulations \cite{Creffield02}
for the pinned XY model and low external magnetic fields. 
However, the possibilty that the transition is not of KTBHNY type was
not excluded \cite{Fisher80}. Indeed, for higher densities of
dislocations (i.e. their core energy is low) a first order transition
was found \cite{Gupta}. Moreover, vortices are not hard core
particles.  This applies in particular if they are described in the
lowest Landau level(LLL) approximation where the density of vortices
is high. With the help of a high
temperature series expansion starting with the LLL approximation
Hikami {\em et al} \cite{Hikami91} found indeed a {\it  first order}
transition from the quasi-long range ordered vortex lattice to the
liquid phase. In this case the intermediate hexatic phase disappears.
The same result was obtained by
Tesanovic and Xing \cite{Tesanovic91} by Monte Carlo simulations
using the LLL approximation \cite{Thouless75,Yoshioka83}.
 However,
theoretical works of Moore and coworkers \cite{Moore92,Yeo96} raise  doubts
about the existence of the quasi-long range ordered
 vortex lattice and suggest that the flux liquid
phase exists at any nonzero temperature.
The simulations based on the LLL approximation
gave conflicting results regarding the
flux lattice melting
in a clean 2D system. Using quasi-periodic boundary condition on a plane,
Hu and MacDonald \cite{Hu}, and Kato and Nagaosa \cite{Kato} have found a
first order transition from a quasi-long range ordered vortex lattice phase
to a vortex liquid. However, the quasi-solid phase was not detected by
simulations on a sphere \cite{ONeil93,Lee94,Dodgson97}.

In the presence of disorder the situation becomes even more complicated.
It has long been predicted that in the presence of frozen-in disorder
any long range crystalline order of a vortex lattice phase is destroyed
in less than four dimensions \cite{Larkin70}. Instead, a liquid-like
phase was expected to occur leaving only short range crystalline order
at length scales smaller than the Larkin length
\cite{Larkin79}. Recently, it was suggested that a Bragg glass phase with
quasi-long range order can exist in three dimensional impure superconductors
\cite{Nattermann90,Korshunov93,Giamarchi94,Emig99}
provided the disorder is weak enough such that the dislocations cannot
proliferate.  Its existence in three dimensions was supported by further
analytical \cite{Kierfeld97,Carpentier96,Fisher97}
and numerical
\cite{Gingras96,Otterlo98} studies.  Experimental evidence
for the existence of the Bragg glass phase was provided recently by  neutron
diffraction data \cite{Klein01}.

The situation in the 2D case is less clear.
For clean 
systems the KTBHNY theory of
melting predicts a dislocation driven melting transition from a positionally
quasi-long range ordered to a hexatic phase. The transition temperature is
$T_m = \frac{a_{\triangle}^2}{4\pi}
\frac{\mu(\lambda+\mu)}{\lambda+2\mu}$.
Here $\lambda = c_{11} - 2c_{66}$ and $\mu = c_{66}$ denote
the effective Lam\'e coefficients, $a_{\triangle}^2 =
\frac{2}{\sqrt{3}}\frac{\phi_0}{B}$ and $\phi_0 = \frac{hc}{2e}$ denotes the 
flux quantum.
For $T < T_m$ the structure factor 
\begin{eqnarray}
S({\bf k}={\bf G}+{\bf q}) \sim
|{\bf q}|^{2-\eta_{{\bf G}}(T)}
\label{eq:stfactor}
\end{eqnarray}
shows quasi-Bragg peaks with a temperature dependent exponent 
\begin{eqnarray}
\eta_{{\bf G}}(T) = \frac{|{\bf G}|^2T}{4\pi\mu}
\frac{\lambda+3\mu}{\lambda+2\mu}.
\label{eq:eta_Tm}
\end{eqnarray}
In the 2D Abrikosov triangular lattice where the screening length diverges with
vanishing thickness of the film, $\lambda$ becomes scale dependent
$\lambda (L) \sim L^2$, implying 
$\eta_{{\bf G}}(T_m) = \frac{|{\bf G}|^2a_{\triangle}^2}{16\pi ^2}
= \frac{1}{3} \; \; (|{\bf G}|^2 = \frac{16\pi ^2}{3a_{\triangle}^2})$.
 In finite systems of size $L_s$,
$\eta_{{\bf G}}(T_m)$ may be larger because the effective
value of $\lambda(L)$ is still finite. Above the melting transition the
translational long range order of the lattice decays on scales 
$L > \xi \approx \exp \big[\frac{bT_m}{T-T_m}\big]^{\nu}$ with
$\nu \approx 0.3696$ \cite{Young79}. In finite systems one will see a 
slightly larger melting temperature defined by $L_s \approx \xi(T)$. The
hexatic phase shows still quasi-long range order of the bond orientations
which will disappear at a second transition at higher temperatures. 
As mentioned already, the possibility that the transition from the 
crystalline to the liquid phase is first order cannot be excluded.

Adding disorder to a {\it purely elastic} 2D system, i.e. if one suppresses
dislocations by hand, there is at low temperatures a glassy phase with
a structure factor
\begin{eqnarray}
S({\bf k}={\bf G}+{\bf q}) \sim
|{\bf q}|^{2-\tilde{\eta}_{{\bf G}}(T)\ln (|{\bf q}|R_a)}
\label{eq:sf_dis}
\end{eqnarray}
where $\tilde{\eta}_{{\bf G}}(T) \sim \big( \frac{T-T_g}{T_g}\big)^2$
with $T_g$ being the glass temperature and
$R_a$ is the length on which the vortex displacements become of the order of
the lattice spacing $a_{\triangle}$. One should also note that $|{\bf q}|$
has to be smaller than $L_{co}^{-1}$ where $L_{co}$ is
a cross-over length scale to the asymptotic behavior.
The phase characterized by the structure (\ref{eq:sf_dis}) was found by
Carpentier and Le Doussal \cite{Carpentier97} following earlier
work by Cardy and Ostlund \cite{Cardy82}. We will call this phase therefore the
Cardy-Ostlund-Carpentier-Le Doussal (COCD) phase. Contrary to the Bragg glass 
phase in D=3 dimensions this phase does not show infinitely sharp Bragg peaks
for $|{\bf q}| \rightarrow 0$ but the Bragg peaks 
saturate at $qR_a \lesssim e^{-2/\tilde{\eta}_G}$ as can be seen from
(\ref{eq:sf_dis}). Moreover,
this phase {\em is not stable} with respect to dislocations which appear
on a length scale \cite{Nattermann00,Carpentier97,LeDoussal00}
\begin{eqnarray*} 
L_{dis} \sim R_a e^{c_1[\frac{T_m-T}{T}\ln (R/R_a)]^{1/2}}
\end{eqnarray*} 
due to the effect of the disorder.
Here $c_1$ is a numerical constant. Since in the thermodynamic limit
the glass temperature $T_g = 
\frac{3}{2\pi}a_{\triangle}^2\frac{\mu(\lambda+2\mu)}{(\lambda+3\mu)}$ 
is much higher than the melting temperature $T_m$ 
($\frac{T_g}{T_m}\big|_{\tilde{\lambda}_{eff} \rightarrow \infty} = 6)$
one will never observe the glass transition. Thus, for system sizes 
$L_s < L_{dis}$ we have to expect to see essentially a temperature driven
melting transition from a low-$T$ COCD phase to a vortex liquid.

It should be noted that the LLL simulations
\cite{Kienappel97} for disordered systems
fail to see any evidence for the existence of the
COCD phase in even weakly disordered 2D superconductors of
spherical geometry. The same problem but for a plane with the quasi-periodic boundary
condition has not been studied yet.

In this paper we employ the LLL approximation and
quasi-periodic boundary condition on
a plane to study
disordered 2D type-II
superconductors by  Monte Carlo (MC) simulations using the standard
Ginzburg-Landau model where the impurities are introduced via
fluctuations of the local critical temperature. The disorder is
assumed to be weak such that the distance $L_{dis}$ between
dislocations 
is much larger than system sizes studied and hence the
dislocations can be excluded. 

In order to study the  glass phase we monitor the Fourier transform of
the density-density correlation function \cite{Hu,Kato}
\begin{eqnarray}
\chi_{DD}({\bf   k}) =  \int_{{\bf   r}} \int_{{\bf  r}'}
\overline{< | \Psi({\bf  r})|^2 | \Psi({\bf  r}')|^2 >} e ^{i{\bf
k}.({\bf r}-{\bf  r}')} , \label{DD_Fourier}
\end{eqnarray}
where $\Psi({\bf  r}) = |\Psi| e^{i\varphi}$
 denotes the superconducting order parameter
and the overbar the disorder averaging, respectively. The value of
this function at ${\bf  k} = {\bf  G}$ with ${\bf  G}$ being a
reciprocal lattice vector defines the intensity of Bragg peaks.
A closely related correlation function is that of the order parameter for
 translational long range order
defined as follows
\cite{Nattermann00}
\begin{equation}
S({\bf  G},{\bf  r}) \; = \; \overline{< \exp(i{\bf G}.[{\bf
u}({\bf  r})- {\bf  u}(0)]) >} . \label{trinv_OP}
\end{equation}
Here ${\bf  u}({\bf  r})$ denotes the displacement field of the
vortex positions. The latter follow from the condition $\Psi({\bf
r})=0$. It is this correlation function which exhibits quasi-long
range order in the Bragg glass phase.
The structure factor is the Fourier transform of $S({\bf G},{\bf r})$
\begin{eqnarray*}
S({\bf k}) = \int d^2r e^{i{\bf kr}} S({\bf G},{\bf r}).
\end{eqnarray*}

In analogy with spin glass theory \cite{Binder86} one may further consider the
positional glass correlation function \cite{Nattermann00}
\begin{equation}
S_{PG}({\bf  G},{\bf  r}) \; = \; \overline{\big|< \exp(i{\bf
G}.[{\bf  u}({\bf  r} )- {\bf  u}(0)]) > \big|^2} \label{pgc_OP}
\end{equation}
which may give signature of the existence of some residual
'glassy' order of the Abrikosov lattice. If $S_{PG}({\bf G},{\bf
r})$ decays not faster than a power law for $|{\bf  r}|
\rightarrow \infty$ then a system is said to be in the positional
vortex glass phase.

In analogy to the positional glass correlation function
 one can define the  gauge-invariant phase-coherent vortex glass
correlation function \cite{FisherFisherHuse91} $C_{VG}({\bf  r})$:
\begin{equation}
C_{VG}({\bf  r}) \; = \; \overline{ |<\Psi({\bf  r}) \Psi^*(0)
e^{i(2\pi /\Phi_0) \int_{\Gamma}d{\bf  r}.{\bf  A}}>|^2} \; . \label{vg_OP}
\end{equation}
Note that $C_{VG}({\bf  r})$ itself depends on the path $\Gamma$
between ${\bf  r}$ and 0 along which the vector potential is
integrated \cite{Nattermann00}.  To make this correlation function
path independent and simultaneously gauge invariant, Moore
\cite{Moore92} proposed to keep only the longitudinal component of
the vector potential.
This corresponds in the symmetric gauge
to the complete neglect of the phase factor $\exp\big[i(2\pi
/\Phi_0) \int_{\Gamma}d{\bf  r}.{\bf  A}\big]$ in Eq.
(\ref{vg_OP}):
\begin{equation}
C_{VG}({\bf  r}) \; = \; \overline{ |<\Psi({\bf  r}) \Psi^*(0)>|^2}.
\label{eq:vgop_sim}
\end{equation}
 The correlations measured in (\ref{eq:vgop_sim}) cannot be probed
directly but have been proposed to characterize the glassy state of
the vortex array.  They are expected to be reflected however in the
current voltage relation although a straightforward derivation of this
connection is still missing \cite{FisherFisherHuse91}.  The relation
between the correlations measured by (\ref{trinv_OP}) and the
dissipation due to flux creep is closer \cite{Nattermann00}.  Since we
do not discuss dynamical properties in this article we consider
(\ref{eq:vgop_sim}) and (\ref{trinv_OP}) (and (\ref{pgc_OP}) as two
competing theoretical tools to characterize the glassy state of the
vortex array. An exponential decay of (\ref{eq:vgop_sim}) -- as we find  --
would suggest , according to \cite{FisherFisherHuse91},   the disappearence 
of the glassy state. This is in contrast to what one would conclude from 
the almost power-law like decay of (\ref{trinv_OP}) which suggest some 
remaining glassy COCD-behaviour in the low temperature state.
Contrary to previous works we will focus on the behavior of correlation 
functions defined by Eqs. (\ref{trinv_OP}), (\ref{pgc_OP})
 and (\ref{eq:vgop_sim}).

In agreement with ref. \cite{Hu,Kato} we find that the clean 2D
system displays the quasi-long range ordered vortex lattice phase at
low temperatures. 
This result is in variance with the result of Moore {\em et al}
\cite{ONeil93,Lee94,Dodgson97} who found the vortex liquid phase at any
temperature
using the same model but with different
geometry.

In the disordered case, contrary to Kienappel and Moore \cite{Kienappel97}, 
we showed that the glassy COCD phase exists at low temperatures.
In this phase the translational invariance order parameter
is found to decay as $S({\bf G},{\bf r})\sim e^{-\tilde{\eta}_{{\bf G}}\ln^2r}$
as predicted
theoretically by Carpentier and Le Doussal \cite{Carpentier97}.
However, due to small system sizes, the
possibility that it
decays as a simple power law is not excluded. 
The clarification of this point requires simulations for considerably
larger system sizes.  Our results indicate, however, that quasi-long 
range order of correlations measured by (\ref{pgc_OP}) exist at low
temperatures. 

In the glassy phase the phase coherent vortex glass order parameter
decays as $C_{VG}({\bf r}) \sim \exp(-r/R_c)$, where the correlation length
$R_c$ depends not only on the temperature but also on the disorder. The
stronger the disorder the larger is $R_c$ suggesting that the phase-coherent
vortex glass ordering becomes more and more favourable as the disorder is
enhanced.

The outline of the paper is as follows. In Sec. II we present the LLL
approach in the Landau gauge. The transition from the vortex liquid
to the ordered phase
of the clean system is studied in
Sec. III. The nature of ordering in the same system but this time with disorder
is discussed in Sec. IV. The exponential decay of the phase
coherent vortex glass is also presented in this section. Finally,
in the last section we summarize our main results.


\section{Model}


Our simulation is based on the phenomenological Ginzburg-Landau
model in the approximation of a uniform magnetic field ${\bf  B}$.
This is a reasonable approximation, since the effective screening
length $\lambda_{eff}=2\lambda_L^2/s$ diverges for vanishing film
thickness $s$. Here $\lambda_L$ denotes the bulk screening length.
The quenched  disorder is introduced via the fluctuations of the
local phase transition temperature. The Ginzburg-Landau free
energy of the superconductor is given by
\begin{eqnarray}
F \; \; &=& \; \; F_{cl} + F_{dis} \; , \nonumber \\
F_{cl} \; \; &=& \; \; \int d^2r \; [ \alpha (T) |\Psi|^2 +
\frac{\beta}{2} |\Psi|^4 + \frac{1}{2m} | {\bf D}
\Psi |^2 ] \; , \nonumber\\
F_{dis} \; \; &=& \; \; \int d^2r \; \alpha_0\delta T_c({\bf  r})
|\Psi ({\bf r})|^2 \; . \label{H_GL}
\end{eqnarray}
Here $\Psi$ is the complex order parameter representing the
macroscopic wave function of the superconducting electrons. ${\bf
D}$ denotes the gauge invariant derivative {\bf D}$ = -i\hbar
\nabla - \frac{2e}{c} {\bf  A}$ with ${\bf  A}$ being the vector
potential. ${\bf  B} = \nabla\times {\bf  A}$, $e$ and $m$ are the
charge and mass of the electron, respectively.
In the simulation we will use the the Landau gauge ${\bf
A}=B(0,x,0)$. We may go back to the symmetric gauge ${\bf
A}=\frac{B}{2}(-y,x,0)$ by a simple gauge transformation
$\Psi_L=\Psi_S e^{ixy/l^2}$.
 $\alpha (T) = \alpha_0(T-T_c)$
and $\beta$ is a constant; $\alpha_0, \beta
>0$. $T_c$ denotes the zero-field mean-field transition
temperature. $\delta T_c ({\bf  r})$ is real and Gaussian
distributed with
\begin{eqnarray}
< \delta T_c({\bf  r}) > \; &=& \; 0 \; , \nonumber\\
< \delta T_c ({\bf  r}) \delta T_c ({\bf  r'}) > \; &=& \;
\zeta^2 T_c^2\xi_0^2 \delta_{\xi_0} ({\bf  r} -{\bf  r'}) \; .
\label{eq:Theta}
\end{eqnarray}
$\delta_{\xi_0}(x)$ is a  $\delta-$function of width of the order
of the zero temperature correlation length
$\xi_0=\hbar/\sqrt{2m\alpha_0T_c}$. The typical fluctuations
$\delta T_c(\bf r)$ of the mean-field transition temperature are
then of the order $\delta T_c\cong \zeta T_c$.
It is convenient to collocate here the main characteristics of the model
\begin{eqnarray}
B_{c2}(0) = \frac{\alpha_0T_cmc}{\hbar e}, \; \;  b = \frac{B}{B_{c2}(0)},
\; \; Gi = \big(\frac{m\beta}{2\pi\hbar^2\alpha_0}\big)^2,
\nonumber\\
l = \sqrt{\frac{\hbar c}{2eB}} = \frac{1}{\sqrt{b}} \xi_0
= \frac{a_0}{\sqrt{2\pi}} = \sqrt{\frac{\sqrt{3}}{4\pi}} a_{\triangle}.
\; \; \; \; \;  \; \;
\label{eq:charac}
\end{eqnarray}
Here $B_{c2}(0)$ denotes the upper critical field at $T=0$ and $l$ the magnetic length. The density and the lattice constant of the flux line lattice
are given by $\frac{1}{a_0^2}$ and $a_{\triangle}$, respectively.
$Gi$ is the Ginzburg number at $T=T_c$.

 In the LLL approximation one expands 
the order parameter $\Psi$ in terms of eigenfunctions of the
operator $\alpha+{\bf D}^2/2m$  of the lowest Landau level only.
This restriction is a reasonable approximation provided that
fluctuations in higher Landau level channels can be neglected.
The precise range of applicability of
the LLL
approximation is still under debate \cite{ONeil93,Li99}. However, one
can expect that this approximation is valid
for $B \gtrsim B_{c2}/13$ \cite{Li99}.
Since the basic length scale of the LLL
approximation is given by the magnetic length $l= \sqrt{\hbar
c/2eB}$ it is convenient to measure from now on all lengths in
units of $l$, i.e. we replace everywhere $\frac{x}{l}$ by
$x$ and $\frac{y}{l}$ by $y$.

We apply quasi-periodic boundary
conditions to the order parameter inside a finite system with
lengths $L_x$ and $L_y$ parallel to the $x-$ and $y-$direction,
respectively. $\Psi (y+L_y,x) = \Psi (y,x)$ and $\Psi (y,x+L_x) =
\exp(-iL_xy/l)\Psi (y,x)$.  Such boundary conditions are
necessary to result in a phase change by $2\pi N_{\phi}$ when
orbiting the system. $N_{\phi} = \frac{L_xL_y}{2\pi l^2}$
denotes the number of vortices in
the system. We will denote $L_x/l=3^{1/4}\pi^{1/2}N_x$ and
$L_y/l=2\pi^{1/2}N_y/3^{1/4}$, where $N_x$ and $N_y$ are the
numbers of vortex columns and rows, respectively.
 The order parameter can then be written in the form
\cite{Yoshioka83}
\begin{eqnarray}
\Psi ({\bf  r}) = \sqrt{ \frac{|\alpha _B| l\pi^{\frac{1}{2}}
}{\beta L_y}} \sum_{j=1}^{N_{\phi}} \sum_{s=-\infty}^{\infty}
C_{j} e^{\frac{iy X_{j,s}-(x-X_{j,s})^2}{2} } . \label{wf_LLL}
\end{eqnarray}
In the last equation we have introduced the following notations
\begin{eqnarray}
\alpha _B \; &=& \; \alpha (T) + \hbar eB/m
= \alpha_0T_c(1-b-t), \nonumber\\ t &=&\frac{T}{T_c},
X_{j,s} \; = \; j2\pi l/L_y + sL_x/l
\label{alphaH}
\end{eqnarray}

In the mean field theory the
superconducting transition of clean systems occurs at
$T_c(B)=T_c-\hbar eB/\alpha_0mc$ which is defined from the condition
$\alpha _B =0$. $X_{j,s}$ is the center of the cyclotron motion.
The number of vortices $N_{\phi}$ must be chosen to be an integer,
the coefficients $C_{j}$ are in general complex.

\begin{widetext}
Using the expansion (\ref{wf_LLL}) we can rewrite $F_{cl}$ as follows
\begin{eqnarray}
  \frac{F_{cl}}{\epsilon^2 T} = sign(\alpha_B) \sum_j |C_j|^2 +
  \frac{3^{1/4}}{8\sqrt{2}N_y} \sum_{n_s=-\infty}^{\infty}
  \sum_{n_d=-\infty}^{\infty} \bigg[ f(n_s) \bigg| \exp
  [-\frac{\sqrt{3}\pi n_d^2}{N_{\phi}}] C_{[n_s+n_d]} C_{[n_s-n_d]}
  \bigg|^2+ \nonumber\\ +f(n_s+1/2) \bigg| \exp [-\frac{\sqrt{3}\pi
    (n_d-1/2)^2}{N_{\phi}}] C_{[n_s+n_d]} C_{[n_s-n_d+1]} \bigg|^2
  \bigg] ,
\end{eqnarray}
where $C_{[n_s]}=C_{mod(n_s,N_{\phi})}$. The only model dependent
parameter of the model in the clean limit is $\epsilon $ given by
\begin{equation}
 \epsilon=\frac{\alpha_B\pi^{1/2} l}{\beta^{1/2} T^{1/2}}=
 \frac{1}{2Gi^{\frac{1}{4}}}\frac{(b+t-1)}{b^{1/2}t^{1/2}},
\label{eq:epsilon}
\end{equation}
which measures the distance to the mean-field phase boundary $\epsilon = 0$.
Here $b$ denotes the magnetic field in units of the mean-field
upper  critical field  $B_{c2}$ at $T=0$ (compare (\ref{eq:charac})).
Below we will find the phase transitions at a
critical value $\epsilon_c$ of $\epsilon$ which can be inverted
into a band of transition lines $b_c(t)$ parameterized by the Ginzburg
number of the corresponding superconductor:
\begin{equation}
b_c(t;Gi,\zeta=0) = \tilde{b}_c(t,\epsilon_c Gi^{\frac{1}{4}})
\equiv  1 - t + \varkappa_c^2t - \varkappa_c \sqrt{2t(1-t) + \varkappa_c^2t^2},
\; \varkappa_c = \sqrt{2}\epsilon_cGi^{\frac{1}{4}} .
\label{eq:b_c}
\end{equation}
 Finally,
\begin{equation}
f(n_s)=Erf(N_x 3^{1/4}(2\pi )^{1/2} - n_sN_y^{-1}3^{1/4}(2\pi )^{1/2}) +
Erf(n_sN_y^{-1}3^{1/4}(2\pi )^{1/2}),
\end{equation}
where $Erf(x)$ is an error function.
It should be noted that different groups used different notations for
the dimensionless parameter $\epsilon$. Our $\epsilon$ =
$t$ (Ref. \onlinecite{Hu}) $= t/2$ (Ref. \onlinecite{Kato}) $= 
\alpha_T/\sqrt{2}$ (Ref. \onlinecite{Hikami91}) $= g/\sqrt{2}$
(Ref. \onlinecite{Tesanovic91}).

To express the disorder term $F_{dis}$ in terms of the LLL coefficients $C_j$,
we expand the random Gaussian disorder in renormalized Hermite polynomials
$u_k(x)$ and  harmonics
\begin{equation}
\delta T_c({\bf  r})= \frac{\zeta T_c\xi_0}
{(L_yl)^{1/2}}\sum_{k=0}^{\infty}
\sum_{m=-\infty}^{\infty}a_{km}e^{2\pi i myl/L_y}
u_k(x),\quad
 u_k(x) =(2^k k! \sqrt{\pi})^{-1/2} e^{-x^2/2}H_k(x).
\label{eq:deltaTc}
\end{equation}
Here $H_k(x)$ are Hermite polynomials and $a_{km}$ are complex
random numbers. To satisfy $\delta T_c({\bf  r})^* =\delta
T_c({\bf r})$ one has to choose $a_{km}$ such that $a_{k,-m} =
a_{k,m}$. Using the standard orthogonality relations
one can show that $\delta T_c ({\bf  r})$ satisfies 
Eq. (\ref{eq:Theta}) if
$< a_{km} > = 0$ and  $< a_{km} a_{k'm'}^* > = \delta_{mm'} \delta_{kk'}$.
From Eqs. (\ref{wf_LLL}) and (\ref{eq:deltaTc}) we have
\begin{eqnarray}
  \frac{F_{dis}}{T} = \epsilon^2\tilde{\zeta}(\frac{l}{L_y})^{1/2}
\sum_{k=0}^{\infty}  \sum_{n_s,n_d}\Big( a_{k,2nd} C_{[n_s+n_d]}
C_{[n_s-n_d]}^*  L(k,n_s,n_d) + \nonumber\\ a_{k,2nd-1} C_{[n_s+n_d]}
C_{[n_s-n_d+1]}^* L(k,n_s+\frac{1}{2},n_d-\frac{1}{2}) \Big) .
\end{eqnarray}

\end{widetext}
Here
\begin{equation}
L(k,n_s,n_d)  = e^{ -\frac{\pi 3^{1/2}}{N_y^2} n_d^2}
\int_0^{L_x/l} \, dx u_k(x)
e^{-(x-\frac{3^{1/4}\pi ^{1/2}}{N_y} n_s )^2 }
\end{equation}
and the dimensionless parameter $\tilde{\zeta}$ which controls the
relative disorder strength has the form
\begin{equation}
\tilde{\zeta} = \frac{\zeta}{2\pi^{1/2}Gi^{1/4}\epsilon t^{1/2}} =
\zeta\frac{b^{1/2}}{\pi^{1/2}(1-t-b)}.
\label{eq:zeta_tilde}
\end{equation}

Substituting (\ref{wf_LLL}) into Eq. (\ref{DD_Fourier}) one can express
the Fourier transform of the density-density correlation function in
the following form (again we measure $k^{-1}$ in units of $l$, i.e.
we replace $kl$ by $k$)
\begin{equation}
\chi_{DD}({\bf  k})= \bigg(\frac{N_{\phi}\alpha_B \pi
l^2}{\beta}\bigg)^2 e^{-k^2/2} \overline{ <|\Delta({\bf  k})|^2
>}\; , \label{DD_Fourier1}
\end{equation}
where
\begin{equation}
\Delta({\bf  k}) = \sum_{j} C^*_{[j-n_y+N_{\phi}]} C_j
e^{\frac{\pi lq_x}{L_y}(n_y-2j)} \; .
\end{equation}
In what follow we will use $\overline{ <|\Delta({\bf  k})|^2 >}$
to characterize the Bragg peaks.


Our MC simulations are performed for coefficients ${C_j} \in$ {\bf
C}$^{N_{\phi}}$ which are updated by the standard Metropolis
algorithm. The candicate for the new configuration is generated by
$C_j \rightarrow C_j + \epsilon \Delta C$ where $\Delta C$ is a
complex number which is randomly chosen from the region
$|\textrm{Re}\Delta C| \le 1$ and $|\textrm{Im}\Delta C| \le 1$ in
the complex plane. $\epsilon$ is chosen to be of order 0.1 so that
the acceptance ratio is $0.3 \sim 0.5$. The physical quantities
are measured every $20 \sim  50$ MC steps.

\section{Clean system}

The aim of this section is twofold. First,  we want to check our code
for the clean system ($\tilde{\zeta}=0$) which was studied in detail
\cite{Hu,Kato}. Second, we consider the spatial behavior of the
translational invariance correlation function which
have not been studied previously by simulations.

\begin{figure}[hbt]
\includegraphics[width=0.8\linewidth]{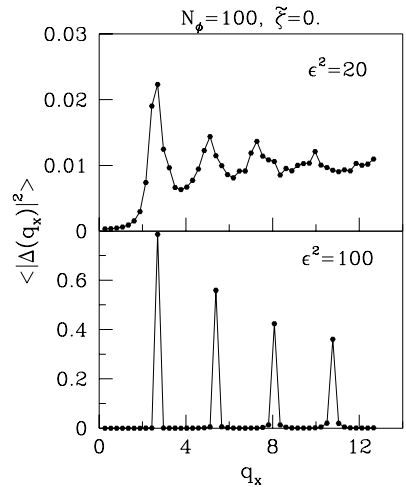}
\caption{The wave vector dependence of the structure factor
for the clean system with $\tilde{\zeta}=0$. $N_{\phi} = 100$,
$\epsilon ^2=20$ and 100. $q_x$ is measured in units $\frac{1}{l}$.}
\label{str_cl_fig}
\end{figure}

The simulations were carried out for systems of five different sizes, each containing $4^2, 6^2, 8^2, 10^2$ and $12^2$ number of vortices. It took
$3\times 10^4 \sim 10^5$ MC steps to reach the thermal equilibration. The
physical quantities were calculated over $10^5 \sim 5\times 10^5$ MC steps.

Fig. \ref{str_cl_fig} shows the temperature dependence of 
$<|\Delta(q_x)|^2> = <|\Delta(q_x,0)|^2>$
for $N_{\phi} = 100$, $\epsilon ^2 = 20$ and 100.
At low temperatures (large $-\epsilon$) this quantity
has the sharp peaks which are indicative
for the existence of the quasi vortex lattice. At
high temperatures the Bragg peaks are smeared out suggesting that we are in
the vortex liquid phase.

\begin{figure}[hbt]
\includegraphics[width=0.8\linewidth]{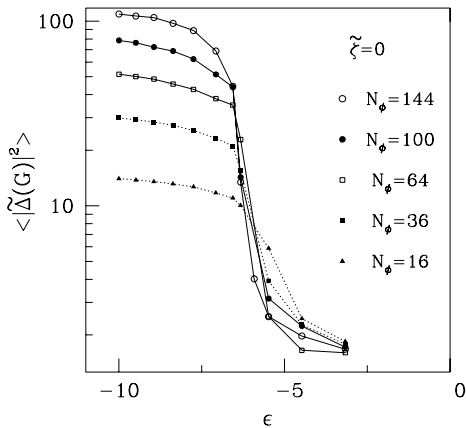}
\caption{The temperature and size dependence of the structure factor
for the clean system. $N_{\phi} = 16, 36, 64, 100$ and 144.}
\label{tem_str0_fig}
\end{figure}

In order to obtain the quasi solid-liquid transition temperature
we monitor the scaling of the renormalized structure factor at
the maximum position $q_x=G=\frac{(4\pi)^{1/2}}{3^{1/4}}$,
$<|\tilde{\Delta}(G)|^2>$, which is defined as follows
\begin{equation}
<|\tilde{\Delta}(G)|^2> \; = \; N_{\phi}
<|\Delta(G)|^2>/<|\Delta(0)|^2> \; .
\label{eq:str_tilde}
\end{equation}
Fig. \ref{tem_str0_fig}
 shows the temperature dependence of $<|\tilde{\Delta}(G)|^2>$
for various values of $N_{\phi}$. The transition temperature is defined
from the point where the curves splay out and we
find $\epsilon ^2_c = 43\pm 2$. This can be compared with previous estimates
$\epsilon ^2_c = 43.5 \pm 1.0$ by Hu and MacDonald \cite{Hu},
$\epsilon ^2_c \sim 50$ by Tesanovic and Xing \cite{Tesanovic91}, 
and $\epsilon ^2_c \approx
49$ by Kato and Nagaosa \cite{Kato}.
The discrepancy between estimates of various groups is probably related to the
fact that the scaling regime was not reached due to small system sizes used
up to now. Since this regime, according to Kato and Nagaosa \cite{Kato},
corresponds to $N_{\phi} > 36^2$ and
the CPU time grows as $N_{\phi}^5$ it is beyond
our computational facilities to reach it.

\begin{figure}[hbt]
\includegraphics[width=0.8\linewidth]{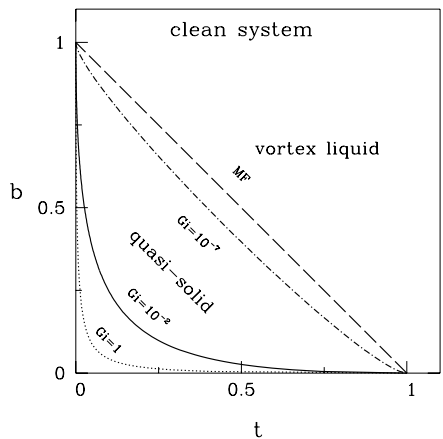}
\caption{The temperature-field phase diagram
for the clean system. The dashed line denotes the mean field boundary
between the vortex liquid and the quasi-solid  phases.
The dot-dashed, solid and dotted lines correspond to the phase
boundary for Ginzburg
number $Gi = 10^{-7}, 10^{-2}$ and 1, respectively.}
\label{diag_cl_fig}
\end{figure}

Fig. \ref{diag_cl_fig} shows the phase boundary between the vortex liquid
and the quasi-solid phase using Eq. (\ref{eq:b_c}). We have
chosen the Ginzburg number $Gi=10^{-7}$ which is typical for low-$T_c$
superconductors, and $Gi= 10^{-2}$ and 1 for high-$T_c$ materials
\cite{Lobb87}. Since the thermal fluctuations enhance with $Gi$,
the quasi-long range ordered vortex lattice shrinks as one 
increases the Ginzburg number. 

\begin{figure}[hbt]
\includegraphics[width=0.8\linewidth]{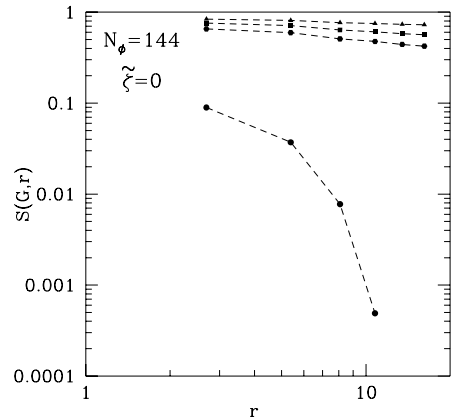}
\caption{The distance dependence of the translational invariance
order parameter $S({\bf  G},{\bf  r})$ for the clean system.
Closed circles, hexagons, squares and triangles correspond to
$\epsilon ^2 = 10, 50, 70$ and 100, respectively. $r$ is measured in
units of $l$.}
\label{trinv_OP_fig}
\end{figure}

\begin{figure}[hbt]
\includegraphics[width=0.8\linewidth]{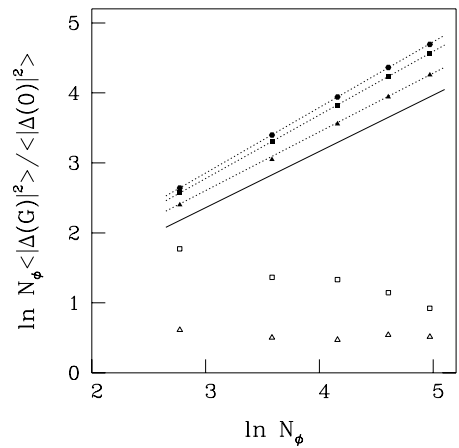}
\caption{Dependence of the structure factor
on $N_{\phi}$ for the clean system.
Open triangles, squares, and closed triangles, squares and hexagons
correspond to
$\epsilon ^2 = 10, 30, 43, 70$ and 100, respectively.  Dotted lines
are linear fits and the solid line corresponds to $\eta_{{\bf G}}=\frac{1}{3}$
from the KTBHNY theory.}
\label{fig:scal_str0}
\end{figure}

\begin{figure}[hbt]
\includegraphics[width=0.8\linewidth]{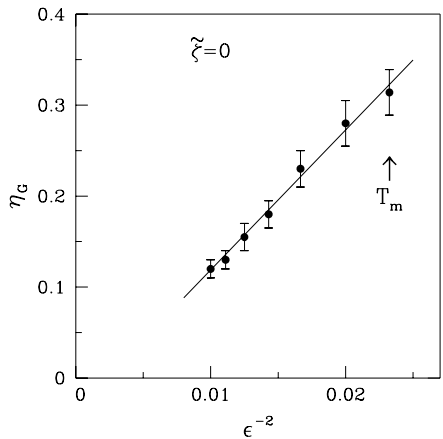}
\caption{ Dependence of
$\eta_{{\bf G}}(T)$ on $\epsilon^{-2}$ for the pure system.
The arrow indicates the melting temperature which corresponds
to $\epsilon^2_c = 43$. }
\label{eta_fig}
\end{figure}

The distance dependence of the translational invariance order parameter
 $S({\bf  G},{\bf  r})$
for $N_{\phi}=144$ and
for various values of $\epsilon$ is shown in Fig. \ref{trinv_OP_fig}.
At low temperatures we find a power law behavior of the correlation
function for translational long range order
\begin{equation}
S({\bf  G},{\bf  r}) \; \sim \; r^{-\eta_{{\bf G}}(T)} \;
\label{S_powerlaw}
\end{equation}
whereas  at high temperatures an exponential dependence on $r$ occurs.

One can define $\eta_{{\bf G}}$ from the distance dependence of 
$S({\bf G},{\bf r})$ shown in Fig. \ref{trinv_OP_fig}. However, in order
to minimize finite size effects we calculate $\eta_{{\bf G}}$ from
the dependence of $<|\tilde{\Delta}(G)|^2>$ on $N_{\phi}$ in the quasi
solid phase:
$<|\tilde{\Delta}(G)|^2> \sim N_{\phi}^{1-\eta_{{\bf G}}/2}$
(see Ref.  \onlinecite{Kato}). In the liquid phase $<|\tilde{\Delta}(G)|^2>$
does not depend on $N_{\phi}$.
Fig. \ref{fig:scal_str0} plots 
ln $<|\tilde{\Delta}(G)|^2>$ versus ln$N_{\phi}$.
The dependence of $\eta_{{\bf G}}(T)$ on parameter $\epsilon$ is presented
in Fig. \ref{eta_fig}. At the melting temperature
$\eta_{{\bf G}}(T_m) =
0.31 \pm 0.03$ which coincides with the value of $\frac{1}{3}$ from the KTBHNY
theory. 
The fact that our results agree with the KTBHNY
predictions suggests that the first order transition is weak and the
second phase transition theory still applies. The weakness of
the first order transition was supported by analytical calculations
of Hikami {\em et al} \cite{Hikami91}
who obtained a small difference between free energies
of the liquid and quasi-long range lattice phases at the melting
temperature.
Since in the LLL approximation screening effects are neglected,
$\lambda \rightarrow \infty$ and hence, as follows from
Eq. (\ref{eq:eta_Tm}), 
$\eta_{{\bf G}} = \frac{G^2T}{4\pi \mu} \sim \frac{1}{\epsilon ^2}$.
The results shown in Fig. \ref{eta_fig} demonstrate that, in agreement
with the theoretical prediction \cite{Nelson79}, 
$\eta_{{\bf G}}(T) \sim \frac{1}{\epsilon ^2}$.

\section{Disordered system}

\begin{figure}[hbt]
\includegraphics[width=0.8\linewidth]{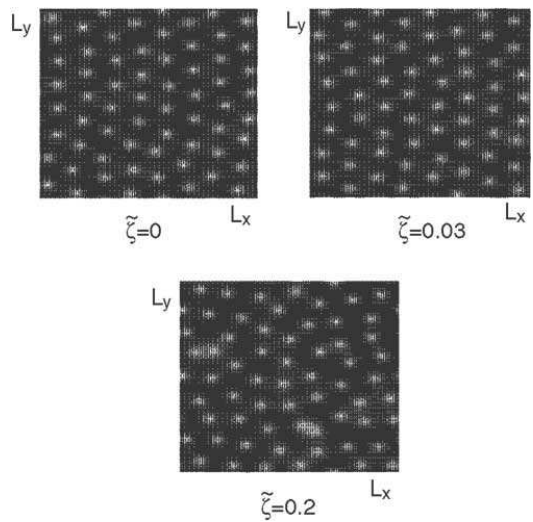}
\caption{Snapshots of the vortex positions for $\tilde{\zeta}=0, 0.03$
and 0.2 obtained at $ \epsilon^2=60$ after 500000 MC steps. Light spots correspond
to vortex cores.
We choose $N_{\phi} = 64$.}
\label{snapshot_fig}
\end{figure}

In this section we will study the effect of weak disorder on the flux lattice
melting transition and on the behavior of different correlation
functions.
Fig. \ref{snapshot_fig} shows a typical snapshots of vortex cores
obtained for the clean case ($\tilde{\zeta} = 0$) and two values of
the disorder strength: $\tilde{\zeta} = 0.03$ and 0.2 after 500000 MC steps.
The runs were done
at $\epsilon^2 = 60$ for the system of 
$N_{\phi}=64$ vortices and are long enough
so that the equilibrium was reached. For the uniform ($\tilde{\zeta}=0$) and
weakly disordered ($\tilde{\zeta}=0.03$) cases we find the slightly
distorted Abrikosov
lattice. The dislocations are clearly seen in the case of $\tilde{\zeta}=0.2$
and the quasi-solid lattice phase cease to be exist.
Thus, in the weak disorder strength limit dislocations can be neglected. We restrict
ourselves to this limit
and concentrate on two values of disorder $\tilde{\zeta}=0.01$ and 0.03.
Some modest simulations have been performed also
for $\tilde{\zeta}=0.02$ and 0.04.

\vspace{0.2in}
\begin{figure}[hbt]
\includegraphics[width=0.8\linewidth]{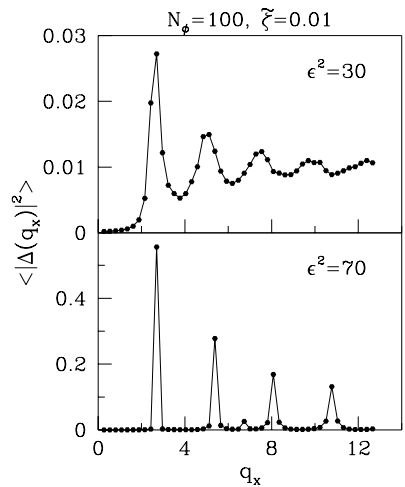}
\caption{The wave vector dependence of the structure factor
for the disordered system with $\tilde{\zeta}=0.01$ .$N_{\phi} = 100,
\epsilon^2=30$ and 70.}
\label{Brpeak_a01_fig}
\end{figure}

The MC simulations were carried out for systems of
$N_{\phi} = 4^2, 6^2, 8^2 10^2$ and $12^2$ vortices. Depending on $N_{\phi}$,
$10^5 - 2\times 10^6$ MC steps were used and half of them were
spent on equilibration of the system. The disorder average is done typically
over $10 - 40$ samples.

Fig. \ref{Brpeak_a01_fig} shows the temperature dependence of the intensity
of Bragg peaks defined by $<|\Delta(q_x)|^2>$ for high ($\epsilon^2 = 30$)
and low ($\epsilon^2 = 70$) temperatures, 
$\tilde{\zeta} = 0.01$ and $N_{\phi} = 100$.
As in the clean case,
at low $T$'s we still have sharp peaks characterizing the 
ordering of the vortices.

\begin{figure}[hbt]
\includegraphics[width=0.8\linewidth]{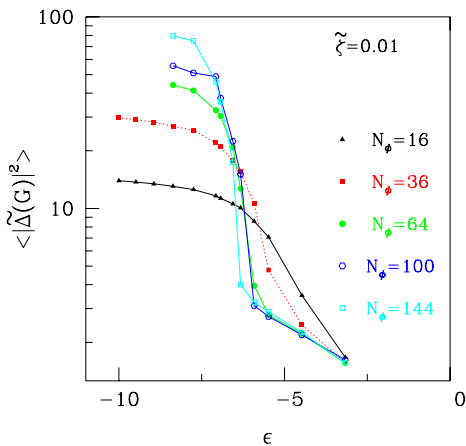}
\caption{The temperature dependence of the structure factor
for the disordered system with $\tilde{\zeta}=0.01$. The curves splay out
at $\epsilon_c^2 = 48 \pm 2$.}
\label{tem_str01_fig}
\end{figure}

The transition temperature to the vortex liquid phase
may be defined from the finite size scaling analysis of the structure
factor. Fig. \ref{tem_str01_fig} shows
the temperature dependence of
$<|\tilde{\Delta}(G)|^2>$ given by Eq. (\ref{eq:str_tilde})
for various system sizes and
$\tilde{\zeta}=0.01$. The curves splay out at $\epsilon_c^2 = 48 \pm 2$ indicating that
the disorder reduces slightly the transition temperature to the liquid phase.
For $\tilde{\zeta} = 0.03$ we have $\epsilon_c^2 = 57 \pm 2$ (the results are not shown).

To calculate from the critical value 
$\epsilon_c(\tilde{\zeta}_j) \equiv \epsilon_{c,j}$ the true phase boundary
$b_c(t,Gi,\zeta)$ for the four given values of $\tilde{\zeta}_j \;
(\tilde{\zeta}_1 = 0.01, \tilde{\zeta}_2 = 0.02,
\tilde{\zeta}_3 = 0.03$ and $\tilde{\zeta}_4 = 0.04$) 
we use the relation (\ref{eq:zeta_tilde}).
Since the maximal value of $t$ is one, the maximum value of $\zeta$ 
is given by $\zeta_j(Gi) = 
2\sqrt{\pi}Gi^{\frac{1}{4}}\tilde{\zeta}_j\epsilon_{c,j}$.
Having chosen for a given Ginzburg number $Gi$ the values of
$\zeta$ smaller than $\zeta_j(Gi)$, we find that the relation
(\ref{eq:zeta_tilde}) defines a temperature $t_j = 
\frac{\zeta^2}{4\pi Gi^{1/2}\tilde{\zeta}^2_j\epsilon^2_{cj}} =
\frac{\zeta ^2}{\zeta^2_{cj}Gi}$
from which we get $b_c(t_j,Gi,\zeta) = 
\tilde{b}_c(t_j,\epsilon_{cj}Gi^{\frac{1}{4}})$. 
Here function 
$\tilde{b}_c$ is defined by Eq. (\ref{eq:b_c}).

\begin{figure}[hbt]
\includegraphics[width=0.8\linewidth]{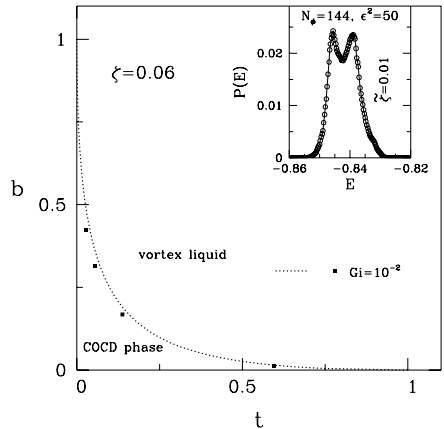}
\caption{The $t-b$ phase diagram
for the disordered system with disorder strength $\zeta=0.06$.
We chose $Gi = 10^{-2}$. The closed squares denote the position of the
melting line in the presence of disorder and the dotted line represents the
case without disorder.
In the disordered case the glassy COCD phase exists at low temperatures.
The inset shows the distribution
$P(E)$ of the
Ginzburg-Landau energy measured in units of the mean-field energy
$E_{MF} = N_{\phi}k_BT\epsilon^2/\beta_A$, where $\beta_A$ is the mean field
value
of the Abrikosov ratio. $P(E)$ was obtained for $N_{\phi}=144$
and $\epsilon^2 = 50$}.
\label{BT_diag_fig}
\end{figure}

Fig. \ref{BT_diag_fig} shows the $b-t$ phase diagram for the disordered case
for $Gi = 0.01$.
The true disorder strength is equal to $\zeta =0.06$. The dashed line 
corresponds to the clean system. Obviously, the fluctuations due to
disorder shift the transition line to lower magnetic fields. 
The inset shows the dependence of
the Ginzburg-Landau energy distribution function $P(E)$ on $E$
for $N_{\phi} = 144$ and
$\epsilon^2 = 50$. The double peak structure gives strong evidence that the
transition is first order.

\begin{figure}[hbt]
\includegraphics[width=0.8\linewidth]{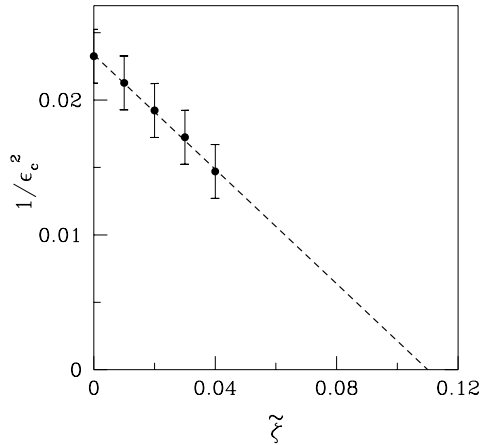}
\caption{The dependence of $\epsilon_c^2$ on $\tilde{\zeta}$. The interpolation
by a straight line
gives $\tilde{\zeta}_c \approx 0.11$.
}
\label{critdis_fig}
\end{figure}

The question we ask now is what is the critical disorder strength
$\tilde{\zeta}_c$ above which the COCD glass phase disappear at any
nonzero  temperature. To answer this question one has to, in principle, find
the dependence of the effective critical temperature $\epsilon_c^2$ on
$\tilde{\zeta}$. But for large values of $\tilde{\zeta}$ it is very
difficult to
equilibrate the system at low temperatures and one cannot locate
 $\epsilon_c^2$.
Therefore, we restrict ourselves to a few values of the disorder
strength and plot $1/\epsilon_c^2$ versus
$\tilde{\zeta}$ as shown in Fig. \ref{critdis_fig}.
The interpolation
to  $1/\epsilon_c^2 = 0$ which corresponds to the real zero temperature,
gives $\tilde{\zeta}_c \approx 0.11$.

\begin{figure}[hbt]
\includegraphics[width=0.8\linewidth]{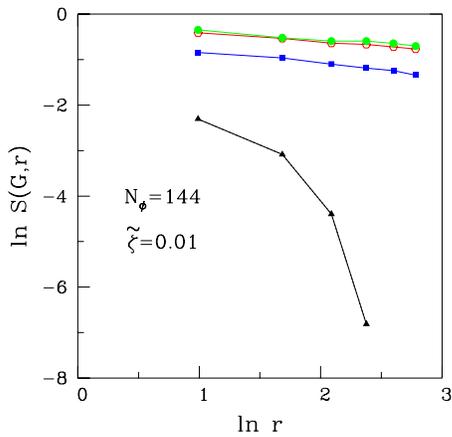}
\caption{The distance dependence of the translational invariance
order parameter $S(G,{\bf  r})$ for the disordered system
with $N$=144 vortices. $\tilde{\zeta}=0.01$, closed triangles,
squares, open hexagons and closed circles correspond to
$\epsilon^2 = 30, 50, 60$ and
70, respectively.
} \label{brag_a01_fig}
\end{figure}

\begin{figure}[hbt]
\includegraphics[width=0.8\linewidth]{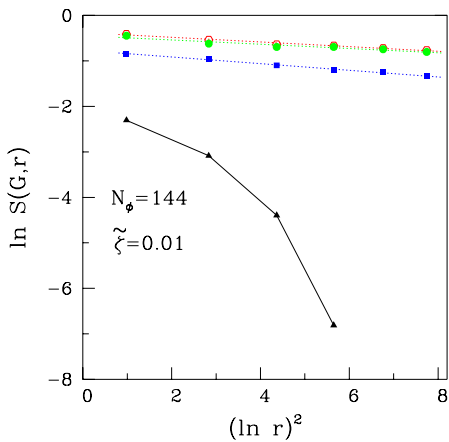}
\caption{The same as in Fig. \protect\ref{brag_a01_fig} but data
are plotted versus ln$^2r$.}
\label{sqlog_a01_fig}
\end{figure}

Fig. \ref{brag_a01_fig} shows the $\ln S(G,{\bf  r}) - \ln r$
plot for $S({\bf G},{\bf  r})$ defined in (\ref{trinv_OP})
of the $N_{\phi} = 12\times 12$ disordered system
with $\tilde{\zeta}=0.01$.
For $\epsilon^2 > \epsilon^2_c = 48$ ln$S({\bf G},{\bf r})$ decays
linearly, except for $\epsilon^2 = 50$ where the decay is much faster.
A slightly better fit can be obtained if one plots ln$S(G,{\bf  r})$
versus ln$^2r$ as shown in Fig. \ref{sqlog_a01_fig}.
In this case $S({\bf G},{\bf  r}) \sim \exp(-\tilde{\eta}_{{\bf G}}\ln^2r)$,
in agreement with the prediction of the COCD theory 
\cite{Carpentier97,Cardy82}.
The exponent $\tilde{\eta}_{{\bf G}}(T)$ depends not only on $T$
but, as we will see below, also on the disorder strength .

\begin{figure}[hbt]
\includegraphics[width=0.8\linewidth]{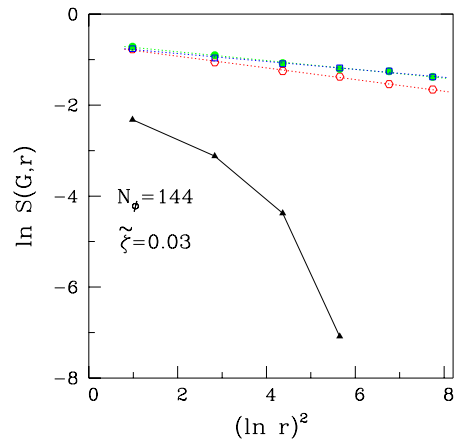}
\caption{The same as in Fig. \protect\ref{sqlog_a01_fig}
 but for $\tilde{\zeta}=0.03$. Closed triangles, open hexagons, closed circles
and open squares correspond to $\epsilon ^2 = 30,60, 70$ and 80,
respectively.}
\label{sqlog_a03_fig}
\end{figure}

Fig. \ref{sqlog_a03_fig} shows the dependence of the translational invariance
order parameter $S(G,{\bf r})$ on $r$ for $\tilde{\zeta} = 0.03$. In this case we still
have the roughness $\sim \ln ^2r$ but the decay is faster than in the
$\tilde{\zeta} = 0.01$ case.

\begin{figure}[hbt]
\includegraphics[width=0.8\linewidth]{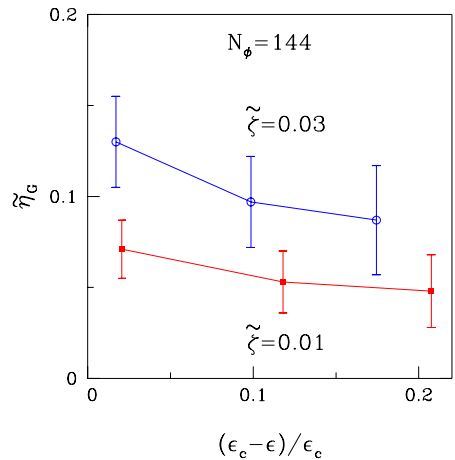}
\vspace{0.2in}
\caption{The temperature and disorder dependence of
$\tilde{\eta}_{{\bf G}}(T)$ characterizing the decay of the translational 
invariance correlation 
function in the super-rough phase.
$ N_{\phi}=144$,
$\tilde{\zeta}=0.01$ (closed squares) and $ \tilde{\zeta}=0.03$ (open
hexagons). Error bars are
mainly due to sample-to-sample fluctuations.}
\label{betaq_fig}
\end{figure}

Fig. \ref{betaq_fig} shows the temperature and disorder dependence of
exponent $\tilde{\eta}_{{\bf G}}(T)$ for $N_{\phi} = 144$.
From the simulation we find that $\tilde{\eta}_{{\bf G}}$ increases with the disorder 
strength. The COCD theory, on the other hand, predicts a universal
disorder independent value of $\tilde{\eta}_{{\bf G}}$, provided one is in the asymptotic
region where one sees true fixed point behavior. We interpret our 
numerical result as a cross-over effect: because of the relatively small
system sizes one is probably not yet in the asymptotic region of the
COCD theory
($L_s \lesssim L_{co}$) and hence an increase of the disorder will results
in a faster decay of the correlation. The other observation is that
$\tilde{\eta}_{{\bf G}}$ increases with increasing temperature. The same remains true
for $\eta_{{\bf G}}$ if we fit
$S({\bf G},{\bf r}) \sim e^{-\eta_{{\bf G}}\ln r}$ (compare Fig. \ref{brag_a01_fig}).
The result for $\tilde{\eta}_{{\bf G}}$ is opposite to the expectation from the
COCD theory in the {\em asymptotic} regime. However, our result can be understood again if we assume that the main effect for the roughness of the vortex
lattice still comes from thermal fluctuations. They alone lead indeed to the increase of $\tilde{\eta}_{{\bf G}}$ with temperature. Thus, we interpret the result of
Fig. \ref{betaq_fig} as a combined effect of thermal fluctuations and
 disorder on scales $L_s \lesssim L_{co}$. The present data does not
 allow for a clear distinction between an ln$r$ and ln$^2r$-behavior
 of ln$S({\bf G},{\bf r})$.

\begin{figure}[hbt]
\includegraphics[width=0.8\linewidth]{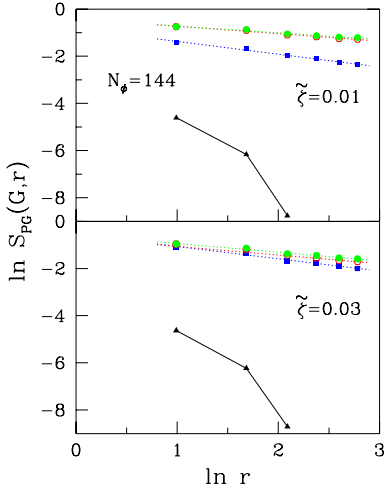}
\vspace{0.2in}
\caption{Ln$S_{PG}(G,{\bf r})-\ln r$ plot for $N_{\phi}=144, \tilde{\zeta}=0.01$
(upper pane)
and $\tilde{\zeta}=0.03$ (lower panel). For $\tilde{\zeta}=0.01$
closed triangles, squares, open hexagons and closed circles correspond
to $\epsilon^2 = 30, 50, 60$ and 70, respectively.
For $\tilde{\zeta}=0.03$
closed triangles, squares, open hexagons and closed circles correspond
to $\epsilon^2 = 30, 60, 70$ and 80, respectively.
}
\label{SPG_log_fig}
\end{figure}

\begin{figure}[hbt]
\includegraphics[width=0.8\linewidth]{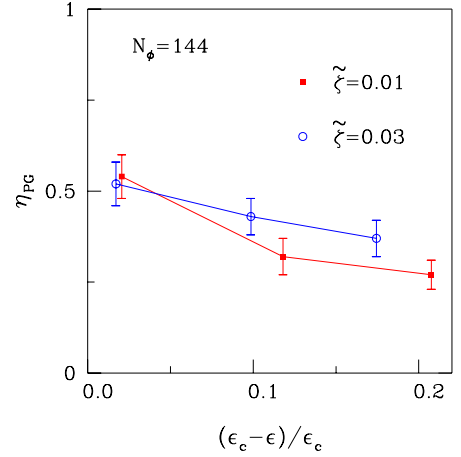}
\vspace{0.2in}
\caption{Temperature dependence of $\eta_{PG}$
for $\tilde{\zeta}=0.01$ (closed squares) and 0.03 (open circles).
}
\label{eta_PG_fig}
\end{figure}

We next study the positional vortex glass ordering characterized by
the order parameter $S_{PG}({\bf  G},{\bf  r})$ (\ref{pgc_OP}) which
we try to fit as
\begin{eqnarray}
\ln S_{PG}({\bf  G},{\bf  r}) \, = \, -\eta_{PG} \ln r + const .
\label{eta12_PG}
\end{eqnarray}
where $\eta_{PG}$ is an exponent. The COCD theory predicts
an ln$ r$-behavior \cite{Goldschmidt82} with $\eta_{PG} \sim T$.
Fig. \ref{SPG_log_fig} shows the distance dependence of $
S_{PG}(G,{\bf  r})$ versus ln$ r$
for $\tilde{\zeta} = 0.01$ and 0.03.
At low temperatures the fit given by Eq. (\ref{eta12_PG})
works well and the corresponding exponent $\eta_{PG}$ is
presented in Fig. \ref{eta_PG_fig}. As the
temperature is lowered the tendency to the 
ordering gets more and more enhanced and $\eta_{PG}$
decreases. Fig. \ref{eta_PG_fig} shows that $\eta_{PG}$ is not sensitive
to the disorder strength.
Both findings are in agreement with the theoretical prediction
\cite{Goldschmidt82} suggesting a quasi-long range glass ordering of
$e^{i{\bf Gu}}$.

\begin{figure}[hbt]
\includegraphics[width=0.8\linewidth]{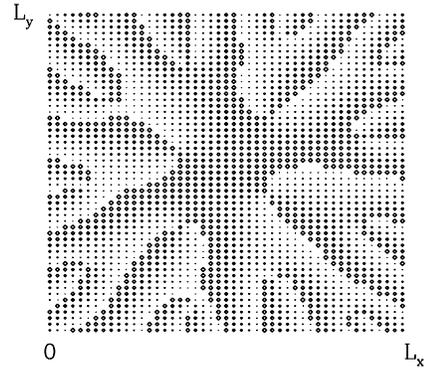}
\caption{Space distribution of the phase of the order
parameter in the symmetric gauge. The darker region the larger
is the value of the phase. The snapshot was obtained for $N_{\phi} = 16$
after 3$\times 10^5$ MC steps at $\epsilon^2 = 50$. The boundary between
the dark and the bright region corresponds to the phase jump from
$\varphi = 2\pi$ to $\varphi = 0$. Going around the system the total phase
change is $2\pi N_{\phi}$.}
\label{Brandt_fig}
\end{figure}

Next we consider the vortex glass correlation function for a phase coherent
vortex glass, Eq. (\ref{eq:vgop_sim}). 
In order to compute the phase-sensitive correlation function $C_{VG}({\bf r})$
we go to the symmetric gauge
to obtain the phase of the order parameter.
The typical spatial distribution of the phase in the symmetric gauge is shown in
Fig. \ref{Brandt_fig}. It has essentially the same shape as presented in the
work of Brandt \cite{Brandt74}.

We fixed the order parameter at the center of the rectangular as shown
in Fig. \ref{Brandt_fig} as $\Psi(0)$ and compute 
$C_{VG}({\bf r})$ with the help of Eq. (\ref{eq:vgop_sim}),
where $|{\bf r}|$ is the distance to this center.
To improve statistics we average over 360 different directions of
{\bf r} keeping $|{\bf r}| = r$ fixed.


\begin{figure}[hbt]
\includegraphics[width=0.8\linewidth]{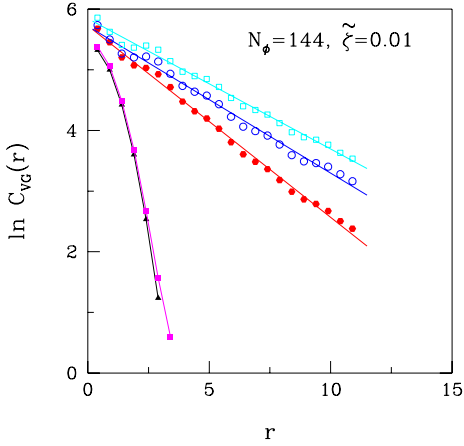}
\caption{ The spatial behavior of the vortex glass order parameter
$C_{VG}({\bf r})$ in the  $\ln C_{VG}({\bf  r})-r$ plot. The closed triangles,
squares, hexagons, open circles and squares correspond to
$\epsilon^2 = 10, 30, 50, 60$ and 70, respectively. Straight lines
are linear fits $\textrm{log} C_{VG}(R) \sim R$ for $\epsilon^2 =50, 60$
and 70.
 $N=144$ and $\tilde{\zeta}=0.01$.}
\label{vgop_12_a01}
\end{figure}

\begin{figure}[hbt]
\includegraphics[width=0.8\linewidth]{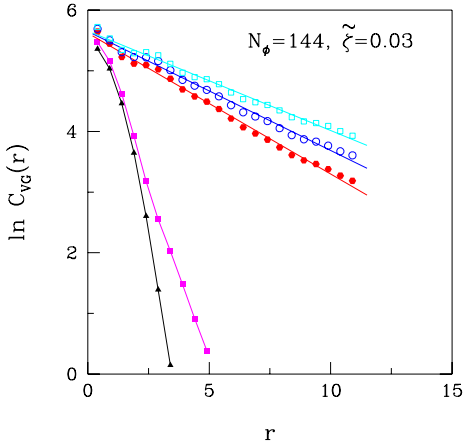}
\caption{ The same as in Fig. \protect\ref{vgop_12_a01}
but for $\tilde{\zeta}=0.03$. 
The closed triangles,
squares, hexagons, open circles and squares correspond to
$\epsilon^2 = 10, 30, 60, 70$ and 80, respectively.
The straight lines are fits for $\epsilon^2 = 60, 70$ and 80.}
\label{vgop_12_a03}
\end{figure}

Fig. \ref{vgop_12_a01} and \ref{vgop_12_a03} show
 the spatial behavior of the vortex glass order parameter
$C_{VG}(r)$ given by Eq. (\ref{eq:vgop_sim})
 for $\tilde{\zeta} = 0.01$ and 0.03.
In the low temperature region straight lines correspond to the fit
\begin{equation}
C_{VG}({\bf r}) \sim \exp(-r/R_c) \, ,
\label{vgop_fit}
\end{equation}
where $R_c$ is the correlation length.
At high temperatures $C_{VG}({\bf r})$ decays faster than exponential.

\begin{figure}[hbt]
\includegraphics[width=0.8\linewidth]{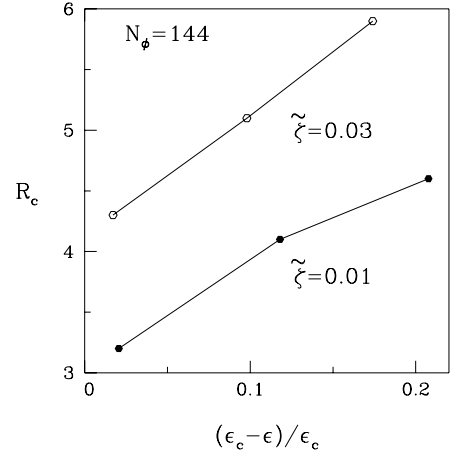}
\vspace{0.2in}
\caption{ Temperature dependence of $R_c$ for the phase coherent
vortex glass order
parameter in the COCD glass phase.
$N_{\phi}=144$, $\tilde{\zeta}=0.01$ and  $\tilde{\zeta}=0.03$.
}
\label{vg_radius}
\end{figure}

Fig. \ref{vg_radius} shows the temperature dependence of $R_c$. 
As the temperature decreases,
the system gets more ordered and $R_c$ increases.
The same tendency shows up if one increases the disorder strength.
However we do not see any sign of a true or quasi-long range order of the
phase coherent vortex glass correlation function.

\section{Conclusion}

We have studied the phase diagram and different correlation functions
in a 2D
type II superconductor by Monte Carlo simulations using the LLL
approximation. For the clean case, in accord with
the previous results of other groups 
\cite{Hikami91,Tesanovic91,Hu,Kato}, we have shown that, 
the quasi-long range ordered vortex lattice phase exists at low temperatures.
The exponent $\eta_{{\bf G}}$ was found to be proportional to
$\epsilon^{-2}$ as expected from the KTBHNY theory of the
dislocation driven melting transition. This result suggests
that the transition from the quasi-solid phase to the liquid one is
weakly first order. It is not excluded that the agreement between our
results and the predictions of the KTBHNY theory is just an artifact
of the finite size effect. 

In disordered systems there is a
first order transition from vortex liquid phase to a COCD
phase characterized by the ln$^2r$-behavior of the roughness.
This phase is however only stable for small enough disorder strength,
$\tilde{\zeta} \lesssim \tilde{\zeta}_c \approx 0.11$. In the COCD
phase the positional vortex glass correlation function decays as power
law. In agreement with the COCD theory, the corresponding exponent 
$\eta_{PG}$ grows with $T$ linearly and remains almost unaffected by the
disorder.

The phase coherent vortex glass correlation function
decays with $r$ exponentially (see Eq. (\ref{vgop_fit})) where the correlation
length $R_c$ increases with disorder indicating that the disorder
favours the vortex glass ordering. However the phase coherent vortex glass
is suppressed by thermal fluctuations in weakly disordered 2D systems.

We thank  Y. Kato and A. Glatz for many useful discussions
and D. Stauffer for his kind help in implementation of the code
for parallel computation
on CRAY located  at the John von Neumann Institute for Computing. The financial support from SFB 608 and the Polish Agency KBN (Grant number 2P03B-146-18)
is acknowledged.


\begin{thebibliography}{10}

\bibitem{Abrikosov57} A. A. Abrikosov, Sov. Phys. JETP {\bf 5}, 1174 (1957).
\bibitem{Eilenberger67} G. Eilenberger, Phys. Rev. {\bf 164}, 628 (1967).
\bibitem{Kosterlitz73} J. M. Kosterlitz and D. J. Thouless, J. Phys. C
{\bf 6}, 1181 (1973).
\bibitem{Berezinsky} V.L. Berezinskii, Sov. Phys. JETP {\bf 32}, 493 (1970),
{\bf 34}, 610 (1971).
\bibitem{Halperin78} B. I. Halperin and D. R. Nelson,  Phys. Rev. Lett.
{\bf 41}, 121 (1978).
\bibitem{Nelson79} D. R. Nelson and B. I. Halperin,  Phys. Rev. B {\bf 19},
2457 (1979)
\bibitem{Young79} A. P. Young, Phys. Rev. B {\bf 19}, 1855 (1979).
\bibitem{Doniach79} S. Doniach and B. A. Huberman,  Phys. Rev. Lett. {\bf 42},
1169 (1979).
\bibitem{Fisher80} D. S. Fisher,  Phys. Rev. B {\bf 22}, 1190 (1980).
\bibitem{Creffield02} C. E Creffield and J. P. Rodriguez, Cond-mat/0205231
\bibitem{Gupta} P. Gupta and S. Teitel, Phys. Rev. {\bf 55}, 2756
(1997) and references therein.
\bibitem{Hikami91} S. Hikami, A. Fujita and A. I. Larkin,
Phys. Rev. B {\bf 44}, 10400 (1991).
\bibitem{Tesanovic91} Z. Tesanovic and L. Xing,  Phys. Rev. Lett. {\bf 67},
2729 (1991)
\bibitem{Thouless75} D. J. Thouless, Phys. Rev. Lett. {\bf 34}, 946 (1975);
G. J. Ruggeri and D. J. Thouless, J. Phys. F {\bf 6}, 2063 (1976).
\bibitem{Yoshioka83} D. Yoshioka, B. I. Halperin,
and P. A. Lee, Phys. Rev. Lett. {\bf 50}, 1219 (1983).
\bibitem{Moore92} M. A. Moore, Phys. Rev. B {\bf 45}, 7336 (1992)
\bibitem{Yeo96} J. Yeo and M. A. Mooore,  Phys. Rev. Lett. {\bf 76}, 1142
(1996);  Phys. Rev. B {\bf 54}, 4218 (1996).
\bibitem{Hu} J. Hu and A. H. MacDonald, Phys. Rev. Lett. {\bf 71}, 432 (1993).
\bibitem{Kato} Y. Kato and N. Nagaosa, Phys. Rev. B {\bf 47}, 2932 (1993);
Phys. Rev. B {\bf 48}, 7383 (1993).
\bibitem{ONeil93} J. A. O'Neill and M. A. Moore, Phys. Rev. B {\bf 48}, 374
(1993)
\bibitem{Lee94} H. H. Lee and M. A. Moore, Phys. Rev. B {\bf 49}, 9240 (1994)
\bibitem{Dodgson97} M. J. W. Dodgson and M. A. Moore,
 Phys. Rev. B {\bf 55}, 3816 (1997).
\bibitem{Larkin70} A. I. Larkin, Sov. Phys. JETP {\bf 31}, 784 (1970).
\bibitem{Larkin79} A. I. Larkin and Yu. N. Ovchinnikov, J. Low. Temp. Phys.
{\bf 34}, 409 (1979).
\bibitem{Nattermann90} T. Nattermann, Phys. Rev. Lett. {\bf 64}, 2454 (1990)
\bibitem{Korshunov93} S. E. Korshunov, Phys. Rev. B {\bf 48}, 3969 (1993)
\bibitem{Giamarchi94} T. Giamarchi and P. Le Doussal, Phys. Rev. Lett. {\bf 72},
1530 (1994);  Phys. Rev. B {\bf 52}, 1242 (1995).
\bibitem{Emig99} T. Emig, S. Bogner and T. Nattermann,
Phys. Rev. Lett. {\bf 83}, 400 (1999); S. Bogner, T. Emig and T. Nattermann,
Phys. Rev. B {\bf 63}, 174501 (2001).
\bibitem{Kierfeld97} J. Kierfeld, T. Nattermann and T. Hwa, Phys. Rev. B
{\bf 55}, 626 (1997)
\bibitem{Carpentier96} D. Carpentier, P. Le Doussal and T. Giamarchi,
Europhys. Lett. {\bf 35}, 379 (1996)
\bibitem{Fisher97} D. S. Fisher, Phys. Rev. Lett. {\bf 78}, 1964 (1997)
\bibitem{Gingras96} M. J. P. Gingras and D. A. Huse,  Phys. Rev. B {\bf 53},
15193 (1996)
\bibitem{Otterlo98} A. van Otterlo, R. T. Scalettar and G. T. Zimanyi,
Phys. Rev. Lett. {\bf 81}, 1497 (1998).
\bibitem{Klein01} T. Klein, J. Joumard, S. Blanchard, J. Marcus, R. Cubitt,
 T. Giamarchi and P. Le Doussal, Nature {\bf 413}, 404 (2001).
\bibitem{Carpentier97} D. Carpentier and P. Le Doussal,  Phys. Rev. B {\bf 55},
12128 (1997).
\bibitem{Cardy82} J. L. Cardy and S. Ostlund, Phys. Rev. B {\bf 25}, 
6899 (1982).
\bibitem{Nattermann00} T. Nattermann and S. Scheidl, Adv. Phys. {\bf 49},
607 (2000).
\bibitem{LeDoussal00} P. Le Doussal and T. Giamarchi, Physica C
{\bf 331} 233 (2000).
\bibitem{Kienappel97} A. K. Kienappel and M. A. Moore, Phys. Rev. B {\bf 56},
8313 (1997).
\bibitem{Binder86} K. Binder and A. P. Young, Rev. Mod. Phys. {\bf 58},
801 (1986).
\bibitem{FisherFisherHuse91} D. S. Fisher, M. P. A. Fisher and D. Huse,
Phys. Rev. B {\bf 43}, 130 (1991).
\bibitem{Li99} D. Li and B. Rosenstein, Phys. Rev. B {\bf 60}, 9704 (1999).
\bibitem{Lobb87} C. J. Lobb, Phys. Rev. B {\bf 36}, 3930 (1987).
\bibitem{Goldschmidt82} Y. Y. Goldschmidt and A. Houghton,
Nucl. Phys. B {\bf 210},
155 (1982)
\bibitem{Brandt74} E. H. Brandt, Phys. Status Solidi {\bf 64}, 257 (1974).
\end{thebibliography}
\end{document}